\documentclass[12pt]{article}
\usepackage[latin9]{inputenc}
\usepackage[a4paper]{geometry}
\usepackage{float}
\usepackage{mathtools}
\usepackage{amsmath}
\usepackage{amsthm}
\usepackage{amssymb}
\usepackage{esint}
\usepackage{pifont}
\usepackage[numbers,sort&compress]{natbib}
\usepackage{color}
\usepackage[unicode=true,
 bookmarks=true,bookmarksnumbered=false,bookmarksopen=false,
 breaklinks=false,pdfborder={0 0 0},pdfborderstyle={},backref=false,colorlinks=false]
 {hyperref}

\def\bal#1\eal{\begin{align}#1\end{align}}
\def\alp[#1]{\begin{align}#1\end{align}}

\def\secnum[#1]{\texorpdfstring{$#1$}{TEXT}}

\def\secnuml#1\secnumr{\texorpdfstring{$#1$}{TEXT}}

\def\eqa{\begin{eqnarray}}
\def\eqae{\end{eqnarray}}
\def\eq{\begin{equation}}
\def\eqe{\end{equation}}
\def\be{\begin{equation}}
\def\ee{\end{equation}}
\def\bea{\begin{eqnarray}}
\def\eea{\end{eqnarray}}
\def\ba{\begin{array}}
\def\ea{\end{array}}
\def\bd{\begin{displaymath}}
\def\ed{\end{displaymath}}

\def\>{\rangle}
\def\<{\langle}

\makeatletter
\numberwithin{equation}{section}
\numberwithin{figure}{section}
\theoremstyle{plain}

\theoremstyle{definition}

\theoremstyle{plain}

\theoremstyle{plain}

\theoremstyle{plain}

\usepackage{braket}
\usepackage{feynmf}
\usepackage{subfigure}

\makeatother

\providecommand{\corollaryname}{Corollary}
\providecommand{\definitionname}{Definition}
\providecommand{\lemmaname}{Lemma}
\providecommand{\propositionname}{Proposition}
\providecommand{\theoremname}{Theorem}

\begin{document}


\begin{titlepage}

\thispagestyle{empty}

\begin{flushright}
\end{flushright}

\vspace{.4cm}
\begin{center}
\noindent{\large \bf  Defect extremal surface as the holographic counterpart of Island formula}\\
\vspace{2cm}

Feiyu Deng$^1$, Jinwei Chu$^{13}$, Yang Zhou$^{12}$
\vspace{1cm}

{\it
$^1$ Department of Physics,
Fudan University, Shanghai 200433, China\\
$^2$ Peng Huanwu Center for Fundamental Theory, Hefei, Anhui 230026, China\\
$^3$ Department of Physics, University of Chicago, Chicago, Illinois 60637, USA
}

\end{center}

\vspace{.5cm}
\begin{abstract}
We propose defect extremal surface as the holographic counterpart of boundary quantum extremal surface. The defect extremal surface is defined by minimizing the Ryu-Takayanagi surface corrected by the defect theory. This is particularly interesting when the RT surface crosses or terminates on the defect. In a simple set up of AdS/BCFT, we find that the defect extremal surface formula gives precisely the same results of the boundary quantum extremal surface. We provide a decomposition procedure of an AdS bulk with a defect brane to see clearly how quantum extremal surface formula emerges from a brane world system with gravity glued to a flat space quantum field theory.
\end{abstract}

\end{titlepage}

\setcounter{tocdepth}{3}
{\hypersetup{linkcolor=black}\tableofcontents}

\newpage

\section{Introduction}

Recent progress in resolving black hole information paradox shows hint towards a new understanding of certain black hole interior as part of the Hawking radiation, which was called island. In particular the island formula for the Von Neumann entropy of Hawking radiation is consistent with unitarity. The black hole evaporation process is expected to be unitary, therefore the Von Neumann entropy of the radiation should follow Page curve~\cite{Page:1993wv,Page:2013dx,Hawking:1976ra}, which shows that the entropy first increases and then decreases at so-called Page time. This happens because, treated as entanglement entropy, the entropy of the radiation can not exceed the black hole entropy for a global pure state.

In recent breakthrough works, a Page curve was computed in AdS black hole plus conformal field theory reservoir~\cite{Penington:2019npb,Almheiri:2019psf}. One key step to reproduce Page curve is to employ the island formula~\cite{Almheiri:2019hni} for the fine grained entropy of radiation, which was inspired from the quantum extremal surface (QES) formula in computing holographic entanglement entropy~\cite{RT:RT-formula,HRT:HRT-formula,Faulkner:2013ana,Engelhardt:2014gca}. For recent related works, see \cite{Almheiri:2019yqk,Almheiri:2019qdq,Penington:2019kki,Rozali:2019day,Chen:2019iro,Verlinde:2020upt,
Chen:2020wiq,Gautason:2020tmk,Anegawa:2020ezn,Giddings:2020yes,
Hashimoto:2020cas,Sully:2020pza,Hartman:2020swn,Hollowood:2020cou,
Alishahiha:2020qza,Geng:2020qvw,Zhao:2019nxk,Chen:2019uhq,Almheiri:2019psy,Li:2020ceg,
Chandrasekaran:2020qtn,Bak:2020enw,Bousso:2020kmy,Dong:2020uxp,Balasubramanian:2020jhl,
Chen:2020jvn,Chen:2020tes,Emparan:2020znc,Hartman:2020khs,Murdia:2020iac,VanRaamsdonk:2020tlr,Liu:2020jsv,Langhoff:2020jqa,Balasubramanian:2020xqf,Balasubramanian:2020coy,
Grado-White:2020wlb,Sybesma:2020fxg,Mirbabayi:2020fyk,Chen:2020uac,Chen:2020hmv,Ling:2020laa,Bhattacharya:2020uun,Marolf:2020rpm,Harlow:2020bee,Nomura:2020ska,Hernandez:2020nem,Chen:2020ojn,Kirklin:2020zic,Matsuo:2020ypv,Goto:2020wnk,Hsin:2020mfa,Akal:2020twv,Numasawa:2020sty,Colin-Ellerin:2020mva,Basak:2020aaa,Geng:2020fxl,Caceres:2020jcn}.

Once promoted to be valid for general quantum system glued with gravity region, the QES formula seems very powerful. In $2d$ Jackiw-Teitelboim (JT) gravity plus CFT model, the QES formula can be derived from the Euclidean path integral computation, which is often called replica wormhole calculation~\cite{Almheiri:2019qdq}. However some puzzles such as JT/ensemble relation arise if one takes the replica wormhole solutions seriously~\cite{Penington:2019kki}. It is still quite interesting to ask how we can test QES formula precisely. Similar situation happens in AdS/CFT, the original Engelhardt-Wall proposal for holographic entanglement entropy has not been proved yet. It is generally believed that QES formula is correct for a bulk with a large $c$ matter, because in that case, both the geometric area term and the leading bulk entanglement are well defined, therefore a generalized minimization procedure for entanglement entropy leads to QES. One can also ask whether it is possible to verify bulk QES formula using boundary CFT techniques, such as conformal bootstrap. In any case, it is still important to find different approaches to derive or even improve QES formula.

In this paper we propose a holographic counterpart for a boundary QES formula, which we call defect extremal surface (DES) formula. This was motivated by the fact that the gravity region relevant in computing QES can be localized on a brane. According to the brane world holography, this can be dual to a brane embedded in a one-dimension higher bulk. On the other hand, the quantum field theory (QFT without gravity) region can also be dual to a one-dimension higher bulk if it is holographic. The boundary condition between gravity region and QFT region can have a higher dimensional dual if it is simple enough such as {\it transparent}. Very similar models have been constructed already in the seminal paper by Almheiri, Mahajan, Maldacena and Zhao~\cite{Almheiri:2019hni}. However we stress that there are two major differences here: First, we do not replace the matter in the gravity region by its holographic dual. Rather we consider it as the defect theory on the brane embedded in the bulk. Second, we do not add additional $2d$ gravity action such as JT action on the brane, rather we consider the $2d$ gravity on the brane purely from the reduction of the bulk. To summarize, compared with~\cite{Almheiri:2019hni}, we replace the gravity itself by some part of the bulk instead of the matter in the gravity region.

Our main idea is to treat the brane as a defect in an AdS bulk. In general there is a quantum theory living on the defect, which is often called defect theory. The defect theory should be considered as part of the full bulk theory since it is coupled to the bulk.
We focus on the contribution to the holographic entanglement entropy from the defect theory. This contribution is particularly interesting when the classical Ryu-Takayanagi surface crosses the defect or terminates on the defect, in which case the additional entanglement entropy coming from the defect theory can be computed straightforwardly. Inspired by the quantum extremal surface proposal~\cite{Engelhardt:2014gca}, we propose the entanglement entropy including defect contribution is given by the defect extremal surface,
\begin{equation}\label{DES0}
S_{\text{DES}} = \min _{\Gamma,X}\left\{\text{ext}_{\Gamma,X}\left[{\text{Area}(\Gamma)\over 4G_N}+S_{\text{defect}}[D]\right]\right\}\ , \quad X=\Gamma\cap D\ ,
\end{equation} where $\Gamma$ is co-dimension two surface in AdS and $X$ is the lower dimensional entangling surface given by the intersection of $\Gamma$ and the defect $D$.

From the boundary point of view, we have a gravity region glued to a QFT region without gravity, where we can employ QES formula to compute the entanglement entropy (fine grained entropy). We compare the results from two approaches and find that they agree with each other precisely. We consider this agreement as strong evidence that the DES formula is the holographic dual of the QES in the context of defect AdS/CFT. Notice that a simple example of defect AdS/CFT is AdS/BCFT proposed by Takayanagi~\cite{Takayanagi:2011zk}. Above the original AdS/BCFT, we include QFT matter on the bulk brane and extend the previous RT formula of holographic entanglement entropy by the defect version (\ref{DES0}). We also go beyond the boundary entropy interpretation of certain part of RT surface and treat them generally as area terms in the gravity on the brane.

This paper is organized as follows. In Section 2, we briefly review the setup of AdS/BCFT proposed by Takayanagi~\cite{Takayanagi:2011zk}. In Section 3, we propose a defect extremal surface (DES) formula and apply it to calculate the entropy of intervals on the asymptotical boundary, including the contribution of the defect theory in the AdS bulk. In Section 4, we use QES formula to compute the entropy of the same intervals on the boundary. We compare the result with that obtained by DES in Section 3 and find precise agreement. We conclude and discuss future questions in Section 5.

{\it Note added}: While this paper was completed, ~\cite{Akal:2020twv} appeared in arXiv, which has some overlap with our discussion on the brane world.
\section{Review of AdS/BCFT\label{sec2}}

\begin{figure}[h]
  \centering
  \includegraphics[width=14cm,height=7.5cm]{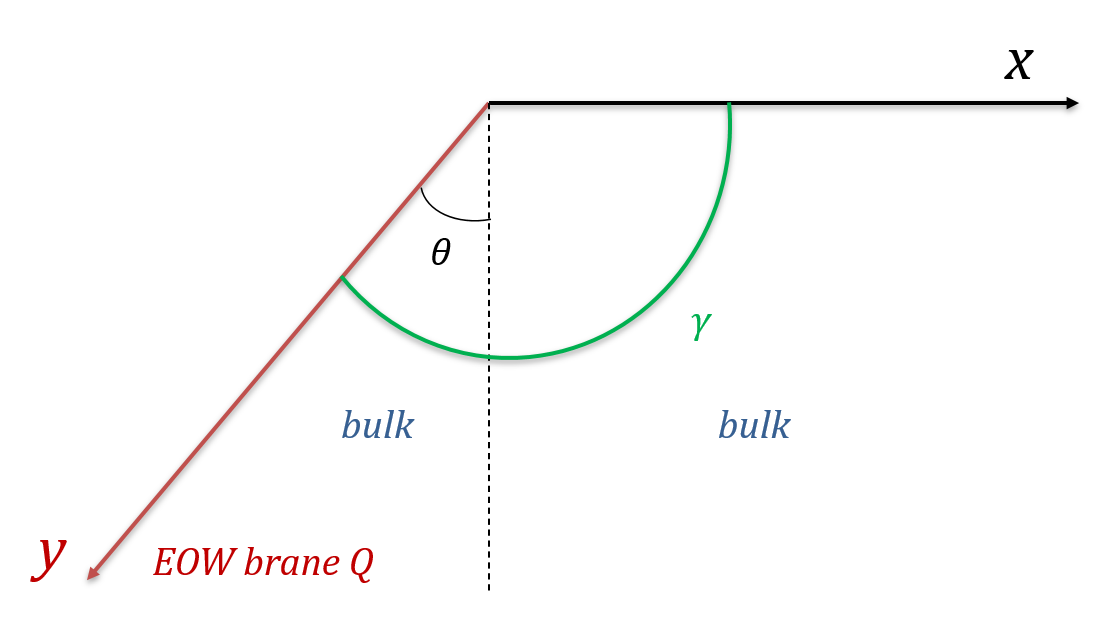}\\
  \caption{Holographic dual of a BCFT$_2$ defined on half space $(x>0)$. }\label{1}
\end{figure}

In this section, we briefly review the setup of AdS/BCFT model proposed by Takayanagi~\cite{Takayanagi:2011zk}.
Consider a 2-dimensional BCFT defined on a half space ($x\ge 0$). The holographic dual of this BCFT in the large central charge limit is a part of a classical AdS$_3$ geometry with a boundary $Q$ where the Neumann boundary condition was imposed. The bulk action is given by~\cite{Takayanagi:2011zk}
\bal
I=\frac{1}{16 \pi G_{N}} \int_{N} \sqrt{-g}(R-2 \Lambda)+\frac{1}{8 \pi G_{N}} \int_{Q} \sqrt{-h}(K-T)\ ,
\eal
where $N$ stands for the bulk and $Q$ the boundary. The Neumann boundary condition demands a tension for the brane $Q$, which is denoted by a constant $T$. The Neumann boundary condition reads
\bal
\label{nbc}
K_{a b}=(K-T) h_{a b}\ ,
\eal
where $h_{a b}$ is the induced metric and $K_{a b}$ is the extrinsic curvature of the $Q$ brane. The metric of AdS$_3$ geometry can be written as
\bal\begin{split}
d s^{2}&=d \rho^{2}+\cosh ^{2} \frac{\rho}{l} \cdot d s_{A d S_{2}}^{2}\\
&=d \rho^{2}+l^2\cosh ^{2} \frac{\rho}{l} \cdot \frac{-d t^{2}+d y^{2}}{y^{2}}\ ,
\end{split}\eal
where $l$ is the AdS radius. The Poincare metric of AdS$_3$ can be recovered by replacing the coordinates $\rho ,y$ with $x,z$
\bal
z=-y / \cosh \frac{\rho}{l}\ , \quad x=y \tanh \frac{\rho}{l}\ .
\eal
If the $Q$ brane is at $\rho=\rho_0$, where $\rho_0$ is a positive constant, it is straightforward to calculate that
 \bal
\label{ec}
K_{a b}=\frac{\tanh \left(\frac{\rho_0}{l}\right)}{l} h_{a b}\ .
\eal
Thus, by combining (\ref{nbc}) with (\ref{ec}), one can determine the relation between the tension and $\rho_0$, i.e.
\bal
T=\frac{\tanh \left(\frac{\rho_0}{l}\right)}{l}\ .
\eal
It is also convenient to choose the polar coordinate $\theta$ with $\frac{1}{\cos (\theta)}=\cosh (\frac{\rho}{l})$. Then, the brane is located at $\theta_0=\arccos (\cosh \frac{\rho_0}{l})>0$, as shown in Fig.\ref{1}.

For an interval $I:=[0,L]$ in BCFT, which contains the boundary, the entanglement entropy can be calculated holographically using (H)RT formula. The minimal surface $\gamma_I$ terminates on the $Q$ brane. The result is
\bal\label{SI}
S_{I}=\frac{\operatorname{Area}\left(\gamma_{I}\right)}{4 G_{N}}=\frac{c}{6} \log \frac{2L}{\epsilon}+\frac{c}{6} \operatorname{arctanh}(\sin \theta_0),
\eal
where $c$ is the CFT central charge and $\epsilon$ is the UV cut off. The RT surface is shown in Fig.\ref{1}. The second term was interpreted as the boundary entropy of BCFT in~\cite{Takayanagi:2011zk}. We will give this term an alternative interpretation in the following.
\section{Bulk defect extremal surface} \label{sec3sideRR}

If the tension of the $Q$ brane is zero, the brane will be orthogonal to the asymptotic boundary, as denoted by the dashed line in Fig.{\ref1}. By adding matter or turn on the tension in the viewpoint of~\cite{Takayanagi:2011zk}, the brane can move to a position with constant $\theta$ angle. We consider the matter on the $Q$ brane as a two-dimensional conformal field theory following~\cite{Almheiri:2019hni} and call it the end of the world (EOW) brane from now on. But we stress that the model considered here is different from that in~\cite{Almheiri:2019hni} in the sense that, the authors there first consider the Dirichlet boundary condition on the boundary brane and then integrated the induced metric respecting to JT gravity. In this work we instead consider Neumann boundary condition on the EOW brane. With CFT matter on the EOW brane, the bulk action is given by
\bal
I=\frac{1}{16 \pi G_{N}} \int_{N} \sqrt{-g}(R-2 \Lambda)+\frac{1}{8 \pi G_{N}} \int_{Q} \sqrt{-h} (K-T) + I_{CFT}\ .
\eal
and the vacuum one point function of the CFT stress tensor is of the form
\bal
\label{st}
\langle T_{ab}\rangle_{\text{AdS}_2}=\chi h_{ab}\ .
\eal
This type of one-point function is reasonable because AdS$_2$ is a maximally symmetric space. The Neumann boundary condition then reads,
\bal
K_{a b}-h_{a b} (K-T)=8\pi G_{N}\chi h_{ab}\ .
\eal
Notice that $\rm\chi$ can be solved from above equation and is also constrained by the trace anomaly on AdS$_2$~\cite{DiFrancesco:1997nk}
\bal
\label{wan}
\langle T^a_{\ a}\rangle=\frac{c'}{24\pi}R\ .
\eal
For the EOW brane located at a constant $\rho_0$, the metric on the brane is determined, we can therefore compute the scalar curvature
\bal
R=-\frac{2}{l^2\cosh^2\frac{\rho_0}{l}}\ .
\eal
Combining the one-point function and the trace formula, one can get the relation between the brane central charge and the tension. Using the Neumann boundary condition, one can further obtain
\bal\label{ccr}
c'=\frac{3l\cosh^2\frac{\rho_0}{l} }{G_N}(\tanh\frac{\rho_0}{l}-lT)=2\cosh^2\frac{\rho_0}{l}(\tanh\frac{\rho_0}{l}-lT)c\ ,
\eal
where the second equality comes from the fact that the CFT central charge on the asymptotic boundary is related to the bulk Newton constant by $\frac{3l}{2G_N}=c$ \cite{Brown:1986nw}.

\subsection{EE for an interval $[y_1,0]$ on the brane}\label{bt1}

Now we consider an interval $[y_1,0]$ on the brane and calculate the entanglement entropy for the ground state of the brane CFT. Notice that CFT on AdS$_2$ can be mapped to a BCFT in flat space via a Weyl transformation~\cite{Giombi:2020rmc,Almheiri:2019psf}. One can read off the Weyl factor from the induced metric on the brane, $ds^2_{brane}= \Omega^{-2}(y) ds^2_{flat}$, i.e.
\begin{equation}\label{wtro}
\Omega (y)=\left|\frac{y\cos \theta_0}{l}\right|\ .
\end{equation}
To compute the entanglement entropy of an interval $[0,a]$ on the brane, one can use the one-point function of the twist operator $\Phi_n$ inserted at $y=a$~\cite{Almheiri:2019psf}. We first consider a BCFT on an upper half plane $y>0$ with the boundary $y=0$. From conformal invariance one can fix the one-point function of the twist operator,\footnote{From the conformal invariance of $(2y)^d \Phi_{\text {flat}}\to (\frac{2y}{\Omega})^d\Omega^d \Phi_{\text {flat}}=(2y)^d \Phi_{\text {flat}}$, one can fix the form of the one-point function of an arbitrary operator $\Phi_{\text {flat}}$ to be
$\langle \Phi_{\text {flat}}\rangle=\frac{g_{\Phi}}{(2y)^d}$
with $g_{\Phi}$ a constant.}
\bal
\left\langle\Phi_{n}\left(y\right)\right\rangle_{\text{flat}}=\frac{g_n}{\left|2y/\epsilon_y\right|^{d_n}}\ ,
\eal where $d_{n}=\frac{c'}{12}\left(n-\frac{1}{n}\right)$ is the conformal dimension of the twist operator. We also include a UV regulator $\epsilon_y$ for the correlation function.

From the Weyl transformation (\ref{wtro}), one obtains
\bal
\left\langle\Phi_{n}(y)\right\rangle_{Q}={\left|\frac{y\cos\theta_0}{l}\right|}^{d_n}\left\langle \Phi_{n}\left(y\right)\right\rangle_{\text {flat}}=g_n{\left|\frac{\cos\theta_0\epsilon_y}{2l}\right|}^{d_n}\ .
\eal
Finally, the entanglement entropy of an interval $[0,y_1]$ on the brane can be computed from the one-point function with the limit of $n\rightarrow 1$ as follows
\bal\begin{split}
S_{\text{bulk}}([y_1,0]) &=\lim _{n \rightarrow 1} \frac{1}{1-n} \log \left\langle\Phi_{n}\left(y_1\right)\right\rangle_{Q}\\
&=\frac{c'}{6} \log \frac{2l}{\cos\theta_0\epsilon_y}+\log g\ \\
&=\frac{c'}{6} \log \frac{2l}{\cos\theta_0\epsilon_y},
\end{split}\eal
where $\log g\equiv \lim _{n \rightarrow 1} \frac{\log g_n}{1-n}$ is the boundary entropy~\cite{Affleck_1994}. For the boundary without physical degrees of freedom, we set $\log g=0$. From this result we see that the entanglement entropy of an interval including the boundary point on the brane is a constant. In particular, it does not depend on the length of the interval. Similar result on AdS$_2$ space has been obtained in~\cite{Almheiri:2019psf}.
\subsection{EE for an interval $[y_1,y_2]$ on the brane}\label{sec32}

Now we calculate the entanglement entropy of an interval $[y_1,y_2]$ that does not contain the boundary $y=0$ on the brane. We insert twist operators at both $y=y_1$ and $y=y_2$. From the Weyl transformation (\ref{wtro}), the two-point function on the brane is related to that on flat space,
 \bal
 \label{wtro2}
\left\langle\Phi_{n}\left(y_1\right)\bar\Phi_{n}\left(y_2\right)\right\rangle_{Q}={\left|\frac{y_1 y_2\cos\theta_0}{l^2}\right|}^{d_n}\left\langle \Phi_{n}\left(y_1\right)\bar\Phi_{n}\left(y_2\right)\right\rangle_{\text {flat}}\ .
 \eal
To compute the two-point function $\left\langle \Phi_{n}\left(y_1\right)\bar\Phi_{n}\left(y_2\right)\right\rangle_{\text { flat }}$ in half flat space, we notice that there are possibly two channels corresponding to bulk operator product expansion (OPE) and boundary operator product expansion (BOE) respectively~\cite{Sully:2020pza}. Which channel dominates can be determined from the cross ratio \footnote{In general, for a couple of points $z_1\equiv x_1+iy_1$ and $z_2\equiv x_2+iy_2$, the cross ratio in BCFT is defined as $\eta=\frac{|z_1-z_2|^2}{4y_1y_2}$.}
\bal
\eta (y_1,y_2)=\frac{(y_1-y_2)^2}{4y_1y_2}\ .
\eal
When $\eta\rightarrow 0$, namely $\Phi_{n}\left(y_1\right)$ and $\bar\Phi_{n}\left(y_2\right)$ are much closer to each other than to the boundary, the OPE channel dominates so that
\bal
\left\langle\Phi_{n}\left(y_1\right) \bar{\Phi}_{n}\left(y_2\right)\right\rangle_{\text{flat}}
=\frac{\epsilon_y^{2 d_{n}}}{\left(y_{1}-y_{2}\right)^{2 d_{n}}}\ .
\eal
Then, with the two-point function on the brane via (\ref{wtro2}), we have the entanglement entropy
\bal
\label{OPEen}
S([y_1,y_2])=\frac{c'}{6}\log \frac{l^2(y_1-y_2)^2}{y_1y_2\epsilon_y^2\cos^2
\theta_0}\ .
\eal
When $\eta\rightarrow \infty$, namely $\Phi_{n}\left(y_1\right)$ or $\bar\Phi_{n}\left(y_2\right)$ is much closer to the boundary than to the other, the BOE channel dominates so that the two-point function is given by
\bal
\left\langle\Phi_{n}\left(y_1\right) \bar{\Phi}_{n}\left(y_2\right)\right\rangle_{\text{flat}}
=\frac{g^{2(1-n)} \epsilon_y^{2 d_{n}}}{\left(4 y_{1} y_2\right)^{d_{n}}}\ .
\eal
Thus, the two-point function on the brane can be obtained via (\ref{wtro2}) and the entanglement entropy is
\bal\begin{split}
\label{BOEen}
S([y_1,y_2])&=\frac{c'}{3} \log \frac{2l}{\epsilon_y\cos\theta_0}+2 \log g\\
&=\frac{c'}{3} \log \frac{2l}{\epsilon_y\cos\theta_0}\ ,
\end{split}\eal
where we set again $\log g=0$ for the boundary without additional degrees of freedom. Again this result does not depend on the position or the length of the interval.

In the large central charge limit, the valid regime of (\ref{BOEen}) and (\ref{OPEen}) can be loosen to $\eta<\eta_c$ and $\eta>\eta_c$ respectively, where $\eta_c$ is the critical value at which $S([y_1,y_2])$ changes from (\ref{OPEen}) to (\ref{BOEen}). By equating these two formulae, one can find that $\eta_c=1$. Therefore, when $c'\to \infty$, the entropy for the brane interval is
\bal
\label{bt2}
\begin{split}
S([y_1,y_2])=
\begin{cases}\frac{c'}{6}\log \frac{l^2(y_1-y_2)^2}{y_1y_2\epsilon_y^2\cos^2
\theta_0},\quad &\eta<1\\
\frac{c'}{3} \log \frac{2l}{\epsilon_y\cos\theta_0},\quad &\eta>1\ .
\end{cases}
\end{split}
\eal

\subsection{Bulk DES: the proposal}\label{33}
In this subsection we propose a defect extremal surface (DES) formula for holographic entanglement entropy. This is particularly useful if the AdS bulk contains a defect $D$. Generally there is a quantum theory living on the defect, which is often called defect theory. As emphasized before, the defect theory should be considered as part of the full bulk theory since it is coupled to the bulk.
We focus on the contribution to the holographic entanglement entropy from the defect theory. This contribution is particularly interesting when the classical Ryu-Takayanagi surface crosses the defect or terminates on the defect, in which case the additional entanglement entropy coming from the defect theory can be computed straightforwardly. Inspired by the quantum extremal surface proposal~\cite{Engelhardt:2014gca}, we would like to propose the entanglement entropy including defect contribution is given by the defect extremal surface,
\begin{equation}\label{DES}
S_{\text{DES}} = \min _{\Gamma,X}\left\{\text{ext}_{\Gamma,X}\left[{\text{Area}(\Gamma)\over 4G_N}+S_{\text{defect}}[D]\right]\right\}\ , \quad X=\Gamma\cap D\ ,
\end{equation} where $\Gamma$ is co-dimension two surface in AdS and $X$ is the lower dimensional entangling surface given by the intersection of $\Gamma$ and $D$. Notice that in general the defect can have any dimension lower than the AdS bulk, but $X$ should be co-dimension two on the defect. In the present work we only focus on the case that $D$ is the boundary of AdS, but the formula (\ref{DES}) should be understood as the proposal for most general cases.

\subsection{Bulk DES for an interval $[0,L]$}
Now we perform the DES calculation for the holographic entanglement entropy corrected by the brane matter in AdS/BCFT. In this case we can use the entanglement entropy on the brane computed in Section \ref{bt1}. Notice that if there is only one brane in the bulk, then $S_{\text{defect}}$ is given by the brane contribution since the rest part is classical AdS without matter field. Also note that when taking extremization, the shape of $\Gamma$ will always be a part of some circle (geodesics) which intersects with the EOW brane at $A$.

\begin{figure}[h]
  \centering
  \includegraphics[width=14cm,height=7.5cm]{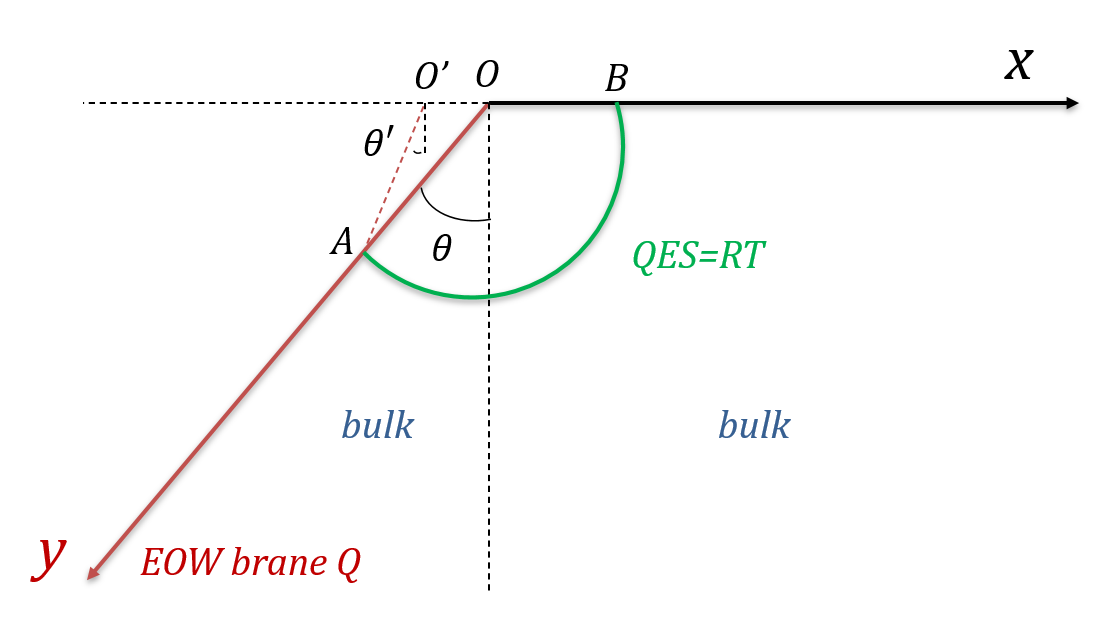}\\
  \caption{DES in original BCFT treated as the RT surface in BCFT$'$. }\label{2}
\end{figure}

Depending on the matter distribution on the brane, the position of $A$ can vary on the brane. In general the new extremal surface following DES is not the previous Ryu-Takayanagi surface. However, there is a convenient way to compute the area term since it is still a part of some circle with center located at asymptotic boundary. One can treat the new surface (after balance) as the RT surface in some new BCFT$'$ as shown in Fig.\ref{2}.

Let $|OB| = L$, $|O'B| = L'$ and $|OA|=a$, one can solve $L'$ and $\theta'$ by simple geometric relations,
\bal
\begin{cases}a \cos\theta=L'\cos\theta'\\
L'-L=-L'\sin\theta'+a\sin\theta\ .
\end{cases}
\eal
The result is
\bal
\begin{cases}L'=\frac{a^2+L^2+2aL\sin\theta}{2(L+a\sin\theta)},\\
\theta'=\text{arcsin}\frac{L^2+2aL\sin \theta-a^2\cos 2\theta}{L^2+2aL\sin \theta+a^2}\ .
\end{cases}
\eal
Then the two terms in (\ref{DES}) can be calculated as
\bal
S_{\text{gen}}(a)=S_{\text{RT}}+S_{\text{defect}}=\frac{c}{6} \log \frac{2L'}{\epsilon}+\frac{c}{6} \operatorname{arctanh}(\sin \theta'_0)+\frac{c'}{6} \log \frac{2l}{\epsilon_y \cos\theta_0}\ ,
\eal
where $\theta'_0$ denotes the value of $\theta'$ when taking $\theta=\theta_0$ and we consider the defect entropy to be the entanglement entropy of the interval $y\in[-a,0]$ on the brane.

By extremizing $S_{\text{gen}}(a)$ with respect to $a$, i.e. $\partial_aS_{\text{gen}}(a)=0$, we get the location of the intersection point between defect extremal surface and EOW brane
\bal
a=L\ ,
\eal
which means that the extremal surface is the same as the RT surface. This is expected because $S_{\text{defect}}$ coming from brane matter is a constant. The final result of the total entanglement entropy is
\bal\label{bgen}
S_{\text{QES}}=\frac{c}{6} \log \frac{2L}{\epsilon}+\frac{c}{6} \operatorname{arctanh}(\sin \theta_0)+\frac{c'}{6} \log \left(\frac{2l}{\epsilon_y \cos\theta_0}\right)\ .
\eal

\subsection{Bulk DES for an interval $[M,L]$ with $M>0$}\label{3.4}
Now we consider an interval $[M,L]$ which does not contain the boundary $x=0$. In this case, we can use the formula derived in Section \ref{sec32}. For simplicity, we will choose to work at $c'=c$.

Note that there are two phases of the defect extremal surface, one is connected and the other is disconnected. We will compute the entropy for the two phases respectively and then compare them.
For the connected phase, the extremal surface does not intersect with the brane. Therefore, there is no $S_{\text{defect}}$ in this case and the entropy is given by the area term, i.e.
\bal\label{cp}
S_{\text{DES}}=\frac{c}{3}\log \frac{(L-M)}{\epsilon}\ .
\eal
\begin{figure}[h]
  \centering
  \includegraphics[width=14cm,height=7.5cm]{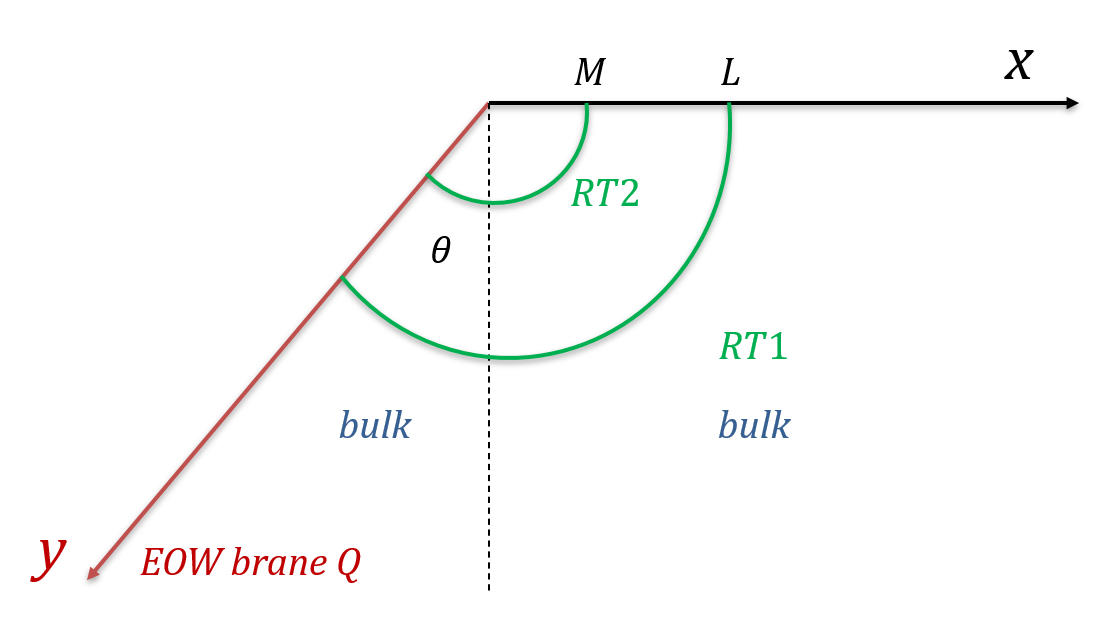}\\
  \caption{Defect extremal surface in the disconnected phase. }\label{3}
\end{figure}
For the disconnected phase, the DES terminates on the brane as shown in Fig.\ref{3} and the entanglement entropy of an interval $[-a,-b]$ on the brane contributes. The generalized entropy includes $S_{\text{defect}}$ given by (\ref{bt2}) (note that the central charge $c'=c$ is large). When the cross ratio on the brane $\eta<1$,
 \bal\begin{split}
S_{\text{gen}}(a,b)=&S_{\text{RT}_1}+S_{\text{RT}_2}+S_{\text{defect}}\\
=&\frac{c}{6}\biggr( \log \frac{a^2 + L^2 + 2 a L \sin\theta_0}{(L + a \sin\theta_0)\epsilon}+\operatorname{arctanh}\frac{L^2+2aL\sin \theta_0-a^2\cos2\theta_0}{L^2+2aL\sin \theta_0+a^2}\\
&+\log \frac{b^2 + M^2 + 2 b M \sin\theta_0}{(M + b \sin\theta_0)\epsilon}+\operatorname{arctanh}\frac{M^2+2bM\sin \theta_0-b^2\cos 2\theta_0}{M^2+2bM\sin \theta_0+b^2}\\
&+\log \frac{l^2(a-b)^2}{ab\epsilon_y^2\cos^2
\theta_0}\biggr)\ .
\end{split}\eal
By extremizing $S_{\text{gen}}(a,b)$ with respect to $a$ and $b$, we find that $\partial_bS_{\text{gen}}(a,b)<0$ for any $a$ and $b$. Thus, there is no extremal solution. When $\eta>1$,
 \bal
 \label{genbt2}
 \begin{split}
S_{\text{gen}}(a,b)=&S_{\text{RT}_1}+S_{\text{RT}_2}+S_{\text{defect}}\\
=&\frac{c}{6}\biggr( \log \frac{a^2 + L^2 + 2 a L \sin\theta_0}{(L + a \sin\theta_0)\epsilon}+\operatorname{arctanh}\frac{L^2+2aL\sin \theta_0-a^2\cos2\theta_0}{L^2+2aL\sin \theta_0+a^2}\\
&+\log \frac{b^2 + M^2 + 2 b M \sin\theta_)}{(M + b \sin\theta_0)\epsilon}+\operatorname{arctanh}\frac{M^2+2bM\sin \theta_0-b^2\cos 2\theta_0}{M^2+2bM\sin \theta_0+b^2}\\
&+2\log \frac{2l}{\epsilon_y\cos\theta_0}\biggr)\ .
\end{split}\eal
By extremizing $S_{\text{gen}}(a,b)$ with respect to $a$ and $b$, i.e. $\partial_aS_{\text{gen}}(a,b)=\partial_bS_{\text{gen}}(a,b)=0$, we get the location of the intersection between defect extremal surface and the EOW brane
\bal
\label{enpb}
\begin{split}
\begin{cases}a=L\\
b=M\ .
\end{cases}
\end{split}
\eal
Following DES proposal
\bal
\label{dcp}
S_{\text{DES}}=S_{\text{gen}}(M,L)=\frac{c}{6}\left( \log \frac{2L}{\epsilon}+ \log \frac{2M}{\epsilon}+2 \operatorname{arctanh}(\sin \theta_0)+2 \log \frac{2l}{\epsilon_y \cos\theta_0}\right)\ .
\eal
Let us compare (\ref{cp}) with (\ref{dcp}). It turns out that the critical point is at $\eta_c(M,L)=e^{2\text{arctanh}\sin\theta_0}(\frac{2l}{\epsilon_y\cos\theta_0})^2$. To summarize,
\begin{equation}
\label{QES1gin}
\begin{split}
S_{\text{DES}}=\begin{cases}\frac{c}{3}\log \frac{(L-M)}{\epsilon},\quad &\eta(M,L)<\eta_c(M,L)\\
\frac{c}{6}\left( \log \frac{4LM}{\epsilon^2}+2 \operatorname{arctanh}(\sin \theta_0)+2 \log \frac{2l}{\epsilon_y \cos\theta_0}\right),\quad &\eta(M,L)>\eta_c(M,L)\ .
\end{cases}
\end{split}
\end{equation}
One can also check that in the disconnected phase, the cross ratio for the endpoints (\ref{enpb}) on the brane satisfies
\begin{equation}
\eta(a,b)=\eta(M,L)>e^{2\text{arctanh}\sin\theta_0}(\frac{2l}{\epsilon_y\cos\theta_0})^2>1\ .
\end{equation}
Hence, the constant phase in (\ref{bt2}) does give the correct defect contribution in (\ref{genbt2}).

\section{Boundary quantum extremal surface\label{sec3}}
In this section we discuss the $2d$ description of the set up in the previous section. A direct way to go from $3d$ to $2d$ is by AdS/CFT correspondence. However in our set up there is  EOW brane in the bulk, which should be treated as part of the bulk. This is because we impose Neumann boundary condition on the brane, which allows the matter on the brane contact with bulk gravity through boundary condition. Note that this is very different from Dirichlet boundary condition~\cite{Cardy:2004hm}. If there is no matter on the brane, in which case the brane is orthogonal to the asymptotic boundary, the $2d$ dual of the bulk is simply a BCFT with zero boundary entropy. If we now turn on the tension of the brane by adding some matter, this would correspond to a non-local deformation for the previous BCFT. Because the matter on the brane is distributed deeply into the bulk, which means that the boundary deformation covers both UV and IR region. So an exact BCFT description of our set up with an nontrivial EOW brane is challenging.

\begin{figure}[h]\label{db}
  \centering
  \includegraphics[width=10cm,height=6cm]{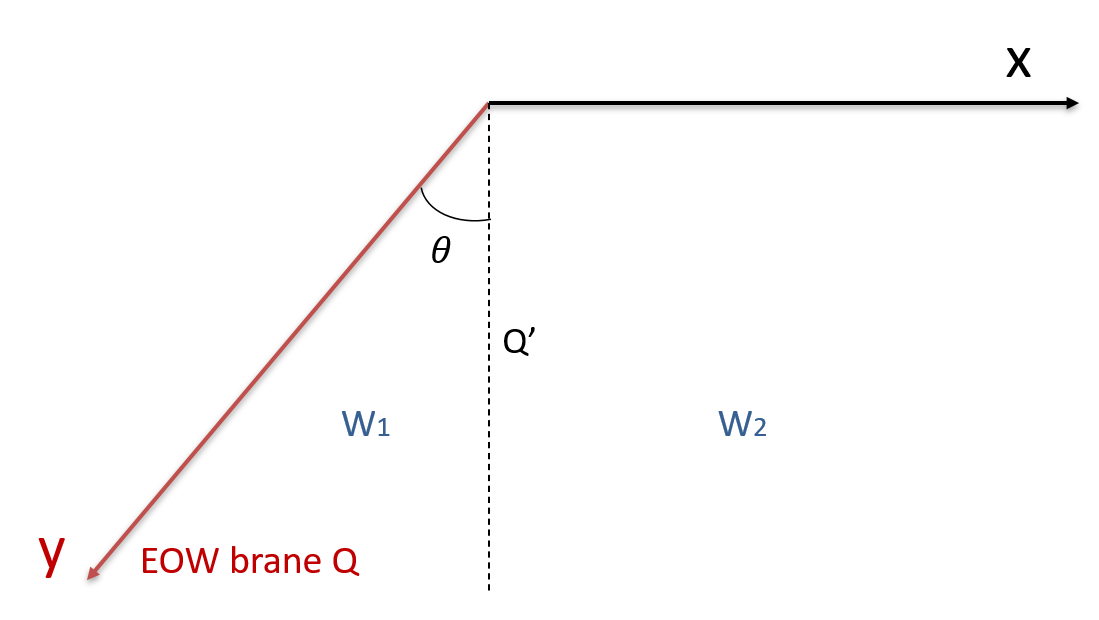}\\
  \caption{Bulk decomposition by inserting an imaginary boundary $Q'$. }\label{A}
\end{figure}

Instead of searching for an exact BCFT description, now we want to find an effective $2d$ description. The ``effective'' here is essentially in the same spirit of semiclassical description of black hole evaporation. The difference is that here we only consider the static situation. To get the $2d$ description for the $3d$ bulk in our set up, we can decompose the bulk into two parts as shown in Fig.\ref{A}. This can be done by inserting an imaginary boundary $Q'$.
Notice that there is no physical degrees of freedom on $Q'$. Now $W_2$ of our bulk is almost the same as the bulk dual of a BCFT with a zero boundary entropy (where the brane is orthogonal to the asymptotic boundary), except for that now the boundary condition on $Q'$ is transparent. Therefore we choose the dual description of $W_2$ in terms of BCFT at the half-space boundary, now with a transparent boundary condition. For $W_1$ we employ the brane world description.

\subsection{Brane world}

\begin{figure}[h]
\hspace{-0.5cm}
  \centering
  \includegraphics[width=15cm,height=4cm]{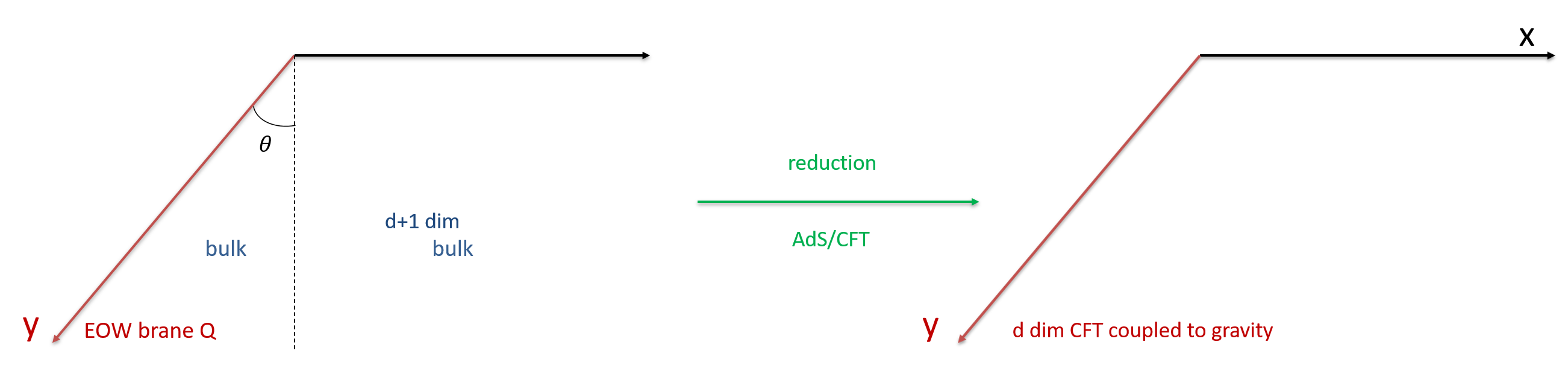}\\
  \caption{Effective description with Neumann boundary condition. }\label{5}
\end{figure}

To find the $2d$ description of $W_1$, one can use the brane world description, i.e. Randall-Sundrum model \cite{Randall:1999vf,Randall:1999ee,Karch:2000ct}. Brane world description is stated as follows. Consider a Poincare AdS$_{d+1}$ with a brane. The Neumann boundary condition is imposed so that on the brane a $d$ dimensional gravity is localized. One can find the effective Newton constant on the brane by doing a Randall-Sundrum reduction along the extra dimension.
In our case the Randall-Sundrum reduction is taken along $\rho$ direction for wedge $W_1$ \cite{Akal:2020wfl}, then the $2d$ gravity theory on $Q$ comes from the reduction of the $3d$ bulk. Together with the brane matter on $Q$, we get the full $2d$ brane theory to be a gravity theory plus CFT on the brane. Now the boundary condition between the brane theory and the half-space CFT should be transparent. This boundary condition is essentially the dual of the bulk boundary condition along $Q'$. Putting everything together, we get the $2d$ effective description as shown in Fig.\ref{5}.

\subsection{Area term}
When we do reduction along $\rho$ direction, $d+1$-dimensional gravity on the wedge is reduced to a $d$-dimensional gravity on the brane and we focus on the curvature term
\bal
\begin{split}
I_{eff}
&\supset\frac{1}{16 \pi G_{N}}\left(\cosh \frac{\rho_0}{l}\right)^{2-d} \int_{0}^{\rho_0} d \rho\left(\cosh \frac{\rho}{l}\right)^{d-2} \int_{\mathrm{Q}} \sqrt{-g^{(d)}} R^{(d)}\\
&\equiv \frac{1}{16 \pi G_{N}^{(d)}} \int_{\mathrm{Q}} \sqrt{-g^{(d)} }R^{(d)}\ ,
\end{split}
\eal
where the superscript $(d)$ denotes the quantities for the effective $d$-dimensional gravity theory on the brane. One can thus find the effective $d$ dimensional Newton constant to be
\bal
\frac{1}{G_{N}^{(d)}}=\frac{1}{G_{N}}\left(\cosh \frac{\rho_0}{l}\right)^{2-d} \int_{0}^{\rho_0} d \rho\left(\cosh \frac{\rho}{l}\right)^{d-2}\ .
\eal
Note that in two dimensions, the area of a point can be determined by the coefficient in front of the Einstein-Hilbert action \cite{Almheiri:2019hni}. Thus the area for a point on the brane in our case is equal to 1, which gives
\bal
\label{art}
\frac{1}{4G_{N}^{(2)}}=\frac{\rho_0}{4G_{N}}=\frac{c}{6}\text{arctanh} (\sin \theta_0)\ .
\eal
This was interpreted as boundary entropy in the original AdS/BCFT proposal~\cite{Takayanagi:2011zk}. Notice that the area term on brane $Q$ is independent of the position and therefore topological.
\subsection{Boundary QES for an interval $[0,L]$}\label{sec43}
Given that the $2d$ effective description contains a brane theory glued with a flat space CFT, now we compute the fine grained entropy from $2d$ perspective by employing the quantum extremal surface (QES) advocated in~\cite{Engelhardt:2014gca} (we verify the QES formula for our set up in Appendix \ref{secrw}). We stress that the $2d$ computation here is independent of holography. For simplicity, we choose to work in the case $c'=c$. We also rescale the flat region coordinates following~\cite{Almheiri:2019qdq} so that the metric becomes $ds^2_{flat}=-dx^+dx^-$, where $x^{\pm}=t\pm x$.

Due to the transparent boundary condition (see Appendix \ref{gc}), the entanglement entropy from CFT can be computed using formula~\cite{Almheiri:2019psf}~\cite{Almheiri:2019qdq}
\bal
S_{\mathrm{CFT}}\left(x_{1}, x_{2}\right)=\frac{c'}{6} \log \left(\frac{\left|x_{1}-x_{2}\right|^{2}}{\epsilon_{1, U V} \epsilon_{2, U V} \Omega\left(x_{1}, \bar{x}_{1}\right) \Omega\left(x_{2}, \bar{x}_{2}\right)}\right)\ ,
\eal
which is the formula for an interval $[x_1,x_2]$ in the metric $d s^{2}=\Omega^{-2} d x d \bar{x}$. The result for the interval $[-a,L]$ in our case is given by
\bal
S_{\text{matter}}([-a,L])=\frac{c}{6}\log \frac{(L+a)^2l}{a\cos\theta_0\epsilon\epsilon_y}\ .
\eal
 Taking into account the area term (\ref{art}), we obtain the generalized entropy,
\bal\begin{split}
S_{\text{gen}}(a)&=S_{\text{area}}(y=-a)+S_{\text{matter}}([-a,L])\\
&=\frac{c}{6} \operatorname{arctanh}(\sin \theta_0)+\frac{c}{6}\log \frac{(L+a)^2l}{a\cos\theta_0\epsilon\epsilon_y}\ .
\end{split}\eal
The extremization condition is $\partial_aS_{\text{gen}}(a)=0$ and the quantum extremal surface is found to be
\bal
a=L\ ,
\eal
which is the same as the end point of the defect extremal surface in Section \ref{33}. The fine grained entropy is
\bal\begin{split}
S_{\text{QES}}&=\frac{c}{6} \operatorname{arctanh}(\sin \theta_0)+\frac{c}{6}\log \frac{4Ll}{\cos\theta_0\epsilon\epsilon_y}\\
&=\frac{c}{6} \log \frac{2L}{\epsilon}+\frac{c}{6} \operatorname{arctanh}(\sin \theta_0)+\frac{c}{6} \log \frac{2l}{\epsilon_y \cos\theta_0}\ .
\end{split}\eal
By comparing the three terms of the above result with (\ref{bgen}), we find that the bulk defect extremal surface result agrees with the boundary quantum extremal surface result precisely.
We consider this precise agreement as a strong support of our DES proposal. This also provides a holographic derivation of the boundary QES formula.
\subsection{Boundary QES for an interval $[M,L]$ with $M>0$}
Now we move to the case for an interval $[M,L]$ which does not contain the boundary $x=0$. Similar to (\ref{3.4}), there are two possible phases in the $2d$ QES computation, one of which contains no contribution from the brane while the other includes the area term as well as the matter entropy from the brane.

Without contribution from the brane, the entropy of $[M,L]$ is just the matter entropy, i.e.
\bal\label{nisp}
S_{\text{QES}}=S_{\text{matter}}([M,L])=\frac{c}{3}\log \frac{(L-M)}{\epsilon}\ .
\eal

Since the brane CFT is coupled to gravity, there is also a possibility that the matter term receives an interval contribution on the brane, denoted as $[-a,-b]$. And the two end points of the interval will also bring area terms, i.e.
\bal
S_{\text{area}}=2\times \frac{1}{4G_N^{(2)}}=\frac{c}{3}\text{arctanh}(\sin\theta_0)\ .
\eal
By employing the entropy formula of two disjoint intervals at large central charge~\cite{Hartman:2013mia} and combining with the area term, the generalized entropy is given by
\bal
\begin{split}
\label{genabBQES}
S_{\text{gen}}(a,b)=&S_{\text{area}}+S_{\text{matter}}\big([-a,-b]\cup [M,L]\big)\\
=&\frac{c}{3}\text{arctanh}(\sin\theta_0)\\
&+\min \left\{ \frac{c}{6}\log \frac{(a-b)^2(L-M)^2l^2}{ab\cos^2 \theta_0 \epsilon^2 \epsilon^2_y}, \frac{c}{6}\log \frac{(L+a)^2(M+b)^2l^2}{ab\cos^2 \theta_0\epsilon^2 \epsilon^2_y} \right\}\ .
\end{split}
\eal
Notice that for the first choice in the ``min", $\partial_bS_{gen}(a,b)<0$, which means that there is no extremal point. If the second choice is picked, under the condition $\frac{(L+a)(M+b)}{(a-b)(L-M)}<1$, the extremization procedure gives
\bal
\begin{split}
\begin{cases}a=L\\
b=M\ .
\end{cases}
\end{split}
\eal
And the final entropy is given by
\bal
\label{isp}
S_{\text{QES}}=S_{\text{gen}}(M,L)=\frac{c}{3}\text{arctanh}(\sin\theta_0)+ \frac{c}{6}\log \frac{16MLl^2}{\cos^2 \theta_0\epsilon^2 \epsilon^2_y}\ .
\eal
By comparing (\ref{nisp}) with (\ref{isp}), one can get the critical point at $\eta=\eta_c$. To summarize,
\begin{equation}
\begin{split}
S_{\text{QES}}=\begin{cases}\frac{c}{3}\log \frac{(L-M)}{\epsilon},\quad &\eta(M,L)<\eta_c(M,L)\\
\frac{c}{6}\left( \log \frac{4LM}{\epsilon^2}+2 \operatorname{arctanh}(\sin \theta_0)+2 \log \frac{2l}{\epsilon_y \cos\theta_0}\right),\quad &\eta(M,L)>\eta_c(M,L)\ ,
\end{cases}
\end{split}
\end{equation}
which is exactly the same as (\ref{QES1gin}). It  can also be checked that in the second phase,
\begin{equation}
\frac{(L+a)(M+b)}{(a-b)(L-M)}=\frac{(2M)(2L)}{(L-M)^2}=\frac{1}{\eta(M,L)}<\frac{1}{\eta_c(M,L)}<1\ ,
\end{equation}
which means that the second choice in the ``min" term of (\ref{genabBQES}) does give the correct matter contribution.
\section{Conclusion and Discussion}\label{sec6}

In this paper we proposed a holographic counterpart of the island formula in the context of defect AdS/CFT. The pioneer work has been done by Almheiri, Mahajan, Maldacena and Zhao~\cite{Almheiri:2019hni} in computing Page curve of the radiation. In the present work we limited ourselves to static case and found a precise holographic derivation of the boundary Island formula. The derivation relies on the bulk {\it defect extremal surface formula} we proposed for holographic entanglement entropy including the contribution from the defect. The derivation also relies on a decomposition of the AdS bulk. From our approach, a $2d$ effective theory including both gravity region and QFT region naturally appears. This is basically because we dualize one part of the bulk by traditional AdS/CFT and do reduction for the remaining part of the bulk using brane world holography. For the $2d$ effective theory, QES formula naturally emerges and we check that the bulk DES and the boundary QES give exactly the same results for different types of single interval entanglement entropy.

Even though we restricted our analysis to the simple case in two dimensions, we expect the results will be similar in higher dimensions.

A few future questions are in order: First, extend our set up to more general cases. So far we focus on the EOW brane located at a constant angle. It would be interesting to generalize the discussion to brane with nontrivial embedding function. A bulk reduction in that case will lead to a different effective $2d$ gravity on the brane. Second, generalize our conjecture DES $=$ QES on boundary to time dependent cases. This can provide a precise holographic dual for a boundary Page curve. In our set up, observables can be computed independently from both the bulk and the boundary, therefore we can use the duality to understand either side physics. Last, one can use our framework to understand the gravity/ensemble relation. Our framework suggests that the ensemble appears in the gravity region because we integrated out part of the bulk. In other words, the $2d$ effective gravity is UV completed by a higher dimensional bulk. We hope to report the progress in future publications.

\section*{Acknowledgements}
We are grateful for useful discussions with our group members in Fudan University. We also thank Rongxin Miao and Jieqiang Wu for useful conversations.
This work is supported by NSFC grant 11905033. YZ is also supported by NSFC 11947301 through Peng Huanwu Center for Fundamental Theory.
\appendix
\section{Boundary QES from replica wormhole}\label{secrw}
As shown in \cite{Almheiri:2019qdq}, QES formula for a system of Jackiw-Teitelboim gravity coupled to CFT can be derived from replica trick, where a replica wormhole needs to be considered as a saddle point. In this appendix, we apply this technique to rederive the boundary QES formula in our set up.

\begin{figure}[h]
  \centering
  \includegraphics[width=8cm,height=6cm]{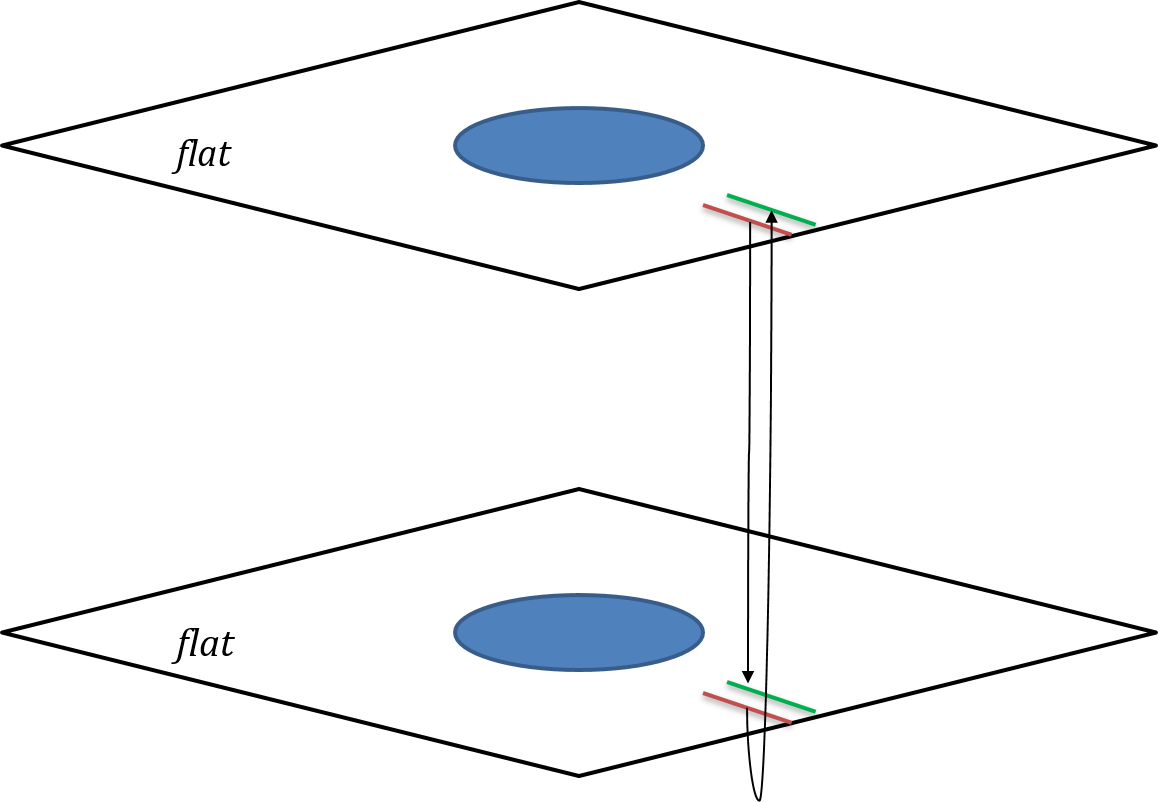}\\
  \caption{Replica trick without replica wormhole for $n=2$ replicas.}\label{7}
\end{figure}

\begin{figure}[h]
  \centering
  \includegraphics[width=8cm,height=6cm]{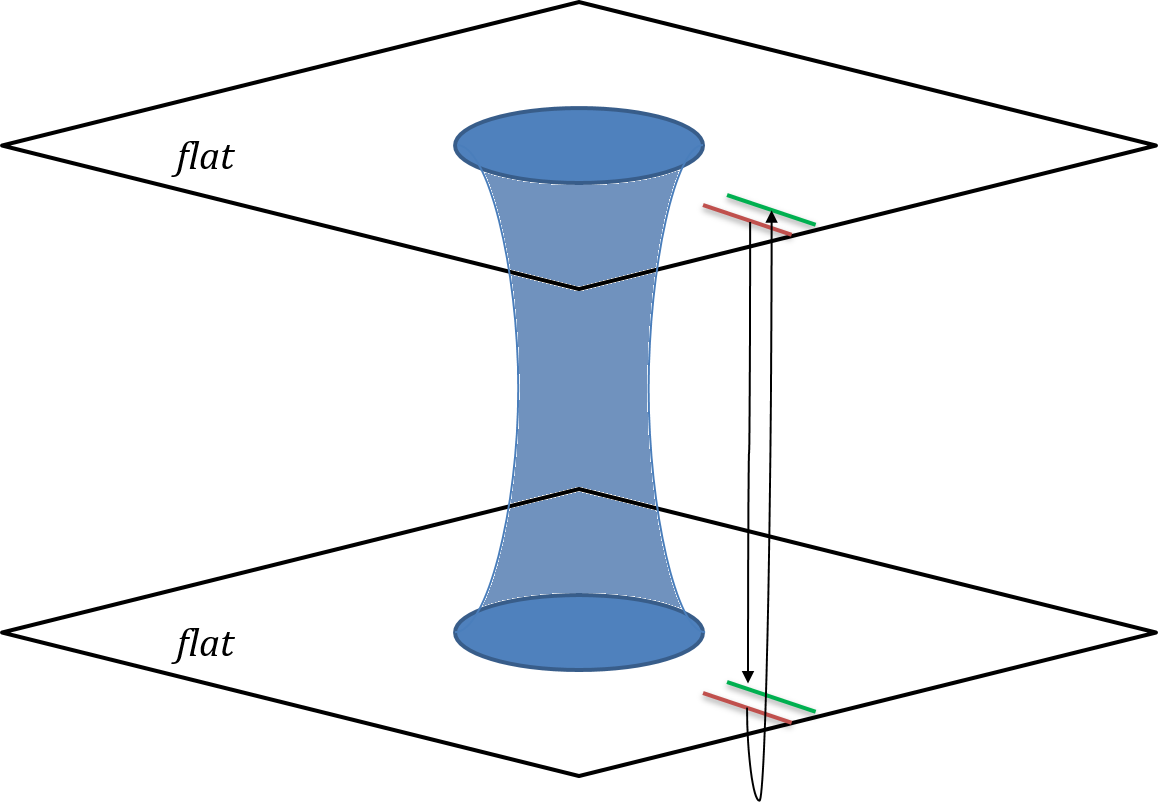}\\
  \caption{Replica wormhole that connects gravity region of the two copies. }\label{8}
\end{figure}

To compute the fine-grained entropy of an interval in the non-gravitational region, we need to use the replica trick to compute the partition function $Z_n$ of a non-trivial manifold (shown in Fig.\ref{7} and Fig.\ref{8} as examples) which contains $n$ copies of the original system. The relationship between the entropy and the partition function can be shown as follows.
\bal
\label{enp}
S=-\partial_n\left(\frac{\log Z_n}{n}\right)\bigg|_{n=1}\ .
\eal
For the non-trivial manifold of replica, there is likely to be many possible saddle points with different topology. One is the Hawking saddle as shown in Fig.\ref{7}, where we simply make a branch cut on the interval in the flat region, glue the $n$-replica cyclically and compute the Euclidean path integral to get $Z_n=\text{Tr}(\rho^n)$. This saddle point corresponds to the usual replica trick for calculating the entanglement entropy of an interval in cases without gravity.
Another saddle point contains a wormhole, which reproduces the entropy from island formula. It permits the gravity region in different copies to dynamically glue with each other, as long as appropriate boundary conditions are satisfied, shown in Fig.\ref{8}.

Now we closely follow~\cite{Almheiri:2019qdq} to derive boundary QES from replica wormhole and will quote some of the conclusions there directly. For readers who are interested in more details we refer to~\cite{Almheiri:2019qdq}.

\begin{figure}[h]
  \centering
  \includegraphics[width=12cm,height=5cm]{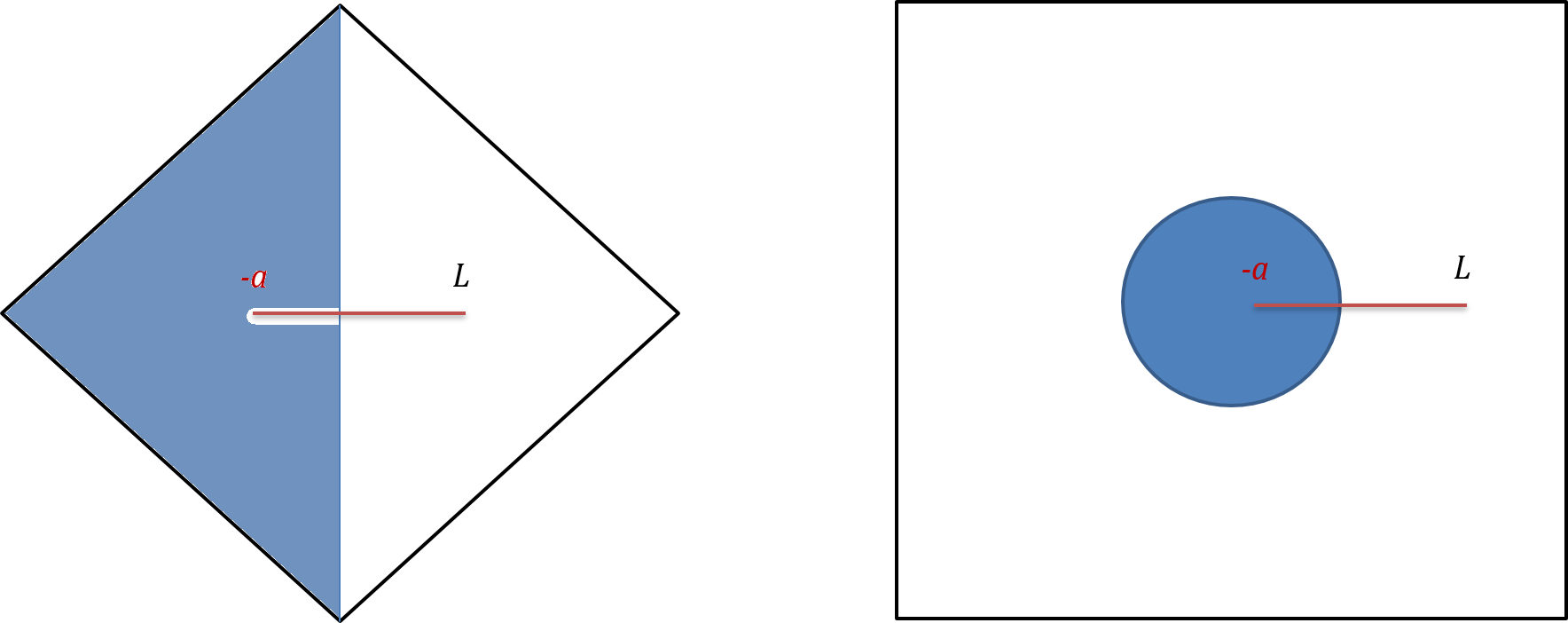}\\
  \caption{$n$-fold cover of effective boundary configuration in Lorentzian signature (left) and Euclidean signature (right). }\label{9}
\end{figure}

For our set up, the total action in effective boundary theory is
\bal
\log Z_{\text{tot}}=\frac{1}{16 \pi G_{N}} \int_{Q} \sqrt{-g }R+\frac{1}{8 \pi G_{N}} \int_{\partial Q} \sqrt{-h}K+\log Z_{\text{CFT}}[g]\ ,
\eal
where the action of the CFT depends on the metric $g$ in the gravity region and is rigid in the flat region. Note that here we write quantities like $G_N^{(2)}$ as $G_N$ for short, not to be confused with 3-dimensional quantities in the context. We will restore them in the last step. Consider an $n$-fold covering surface of this theory, replica wormhole will cause the emergence of an island in the gravity region and we denote the endpoint of the island by $y=-a$ (see Fig.\ref{9}), which is fixed by the saddle point of path integral dynamically. At $y=-a$, there is a conical singularity at which the twist operator of matter field is inserted. After the $Z_n$ quotient, the gravitational action becomes
\bal
-\frac{I_{\text{grav}}}{n}=\frac{1}{16 \pi G_{N}} \int_{Q} \sqrt{-g }R+\frac{1}{8 \pi G_{N}} \int_{\partial Q} \sqrt{-h}K-\left(1-\frac{1}{n}\right)S(\omega),
\eal
where the last term comes from the conical singularity and depends on the position of singularity $\omega$ in general. Here it is just a constant that equals to $\frac{1}{4G_N}$.

We define a complex coordinate $u$ in the gravity region such that the region is given by a disk $|u|\le 1$ with the boundary $u=e^{i \alpha}$ ($\alpha$ is real). Outside the disk is the flat region $x\ge 0$ with $v=e^{i\tilde x}=e^{x+i\tau}$ (where $\tau$ is real) as the coodinate so that $|v| \geq 1$. Then, by using the technique of solving the so called "conformal welding problem" \cite{sharon20062d}, one can find two functions $U$ and $V$ which map the coordinates $u$ and $v$ to another coordinate $z$ holomorphically. More explicitly,
\bal
z=\begin{cases} U(u), \quad |u| \leq 1 \\
 V(v), \quad |v| \geq 1\ .\end{cases}
\eal
It implies that the condition $U\left(e^{i \alpha(\tau)}\right)=V\left(e^{i \tau}\right)$ should be satisfied at the interface $|u|=|v|=1$. Note that the functions $U$ and $V$ are holomorphic inside and outside the disk respectively.

Now we find one equation of motion by varying the total action with respect to $\alpha(\tau)$ since it is a reparametrization mode. Because the gravity action has no dynamical boundary term and the bulk part is topological, $\delta I_{\text{grav}}=0$. Combining it with $\delta I_{\text{CFT}}=-i \int d \tau\left(T_{\tilde x\tilde x}-T_{\bar{\tilde x}\bar{\tilde x}}\right) \frac{\delta \alpha}{\alpha'}$ \cite{Almheiri:2019qdq}, we can thus determine the equation of motion to be
\bal
\label{eom1}
T_{\tilde x\tilde x}=T_{\bar{\tilde x}\bar{\tilde x}},
\eal
where the stress tensors are for one copy of the CFT.

In the $z$ plane, the insertion of twist operators will give non-trivial stress tensors $T_{zz}(z)$ and $T_{\bar z\bar z}(\bar z)$. Then $T_{\tilde x\tilde x}$ and $T_{\bar{\tilde x}\bar{\tilde x}}$ can be obtained from $T_{zz}(z)$ and $T_{\bar z\bar z}(\bar z)$ through the conformal anomaly
\bal
\label{txz}
T_{\tilde x\tilde x}=\left(\frac{d V\left(e^{i \tilde x}\right)}{d \tilde x}\right)^{2} T_{z z}-\frac{c}{24 \pi}\left\{V\left(e^{i \tilde x}\right), \tilde x\right\},
\eal
and a conjugate expression for $T_{\bar{\tilde x}\bar{\tilde x}}$.

Assume that the functions $U$ and $V$ map the two branch points $u_*=e^{-a}$ and $v_*=e^{L}$ to $z=0$ and $z=\infty$ respectively. Then, in a new coordinate $\tilde{z}=z^{1 / n}$, there is no branch point so that the stress tensor is zero. Thus,
\bal
T_{z z}(z)=-\frac{c}{24 \pi}\left\{z^{1 / n}, z\right\}=-\frac{c}{48 \pi}\left(1-\frac{1}{n^{2}}\right) \frac{1}{z^{2}}.
\eal
Then from equation (\ref{txz}), we can get $T_{\tilde x\tilde x}$ and $T_{\bar{\tilde x}\bar{\tilde x}}$ explicitly and finally the equation of motion (\ref{eom1}) becomes
\bal
\label{eom11}
\begin{split}
\frac{1}{2}\left(1-\frac{1}{n^{2}}\right) \frac{V^{\prime}\left(e^{i \tau}\right)^{2}}{V\left(e^{i \tau}\right)^{2}}e^{2i\tau}+\left\{V\left(e^{i \tau}\right), e^{i \tau}\right\}e^{2i\tau}-c.c.=0\ .
\end{split}\eal
As the map $V$ depends on the gluing function $\alpha(\tau)$, this equation is hard to solve generally. However, one can solve it in $n\rightarrow1$ limit as follows.

When $n=1$, the first term in (\ref{eom11}) (as well as its complex conjugate) vanishes. Therefore, the second term is also equal to zero, which means that $V$ is an $SL(2,C)$ transformation. Then, we can construct the functions as follows.
\bal
\label{vn1}
U(u)=V(v)=\frac{v-u_*}{v_*-v},\ u=v\ .
\eal
It can be checked that this function does map two branch points $u_*$ and $v_*$ to $z=0$ and $z=\infty$.

Then, we go near $n=1$ and equate terms of order $n-1$. For the first term in (\ref{eom11}), the $n-1$ term is just that with (\ref{vn1}) plugged in because of the overall $n-1$ coefficient. The second term does not vanish, but the term of order $n-1$ is zero after the Fourier transformation with the mode $k=1$ which switches the coordinate $\tau\to k$. Hence, the Fourier transformation with $k=1$ of the first term (plus its complex conjugate) should also be zero, more explicitly,
\bal
\begin{split}
0&=\int_0^{2\pi}d\tau e^{-i\tau} \left(\frac{V^{\prime}\left(e^{i \tau}\right)^{2}}{V\left(e^{i \tau}\right)^{2}}e^{2i\tau}-c.c\right)\\
&=\int_0^{2\pi}d\tau\left(\frac{e^{i\tau}(u_*-v_*)}{(e^{i\tau}-u_*)^2(e^{i\tau}-v_*)^2}-\frac{e^{i\tau}(u_*-v_*)}{(u_*e^{i\tau}-1)^2(v_*e^{i\tau}-1)^2}\right)\\
&=4\pi \frac{u_*v_*-1}{(u_*-v_*)^2}\ .
\end{split}
\eal
The solution is
\bal
\begin{split}
u_*=\frac{1}{v_*}\ ,
\end{split}
\eal
namely $a=L$, which gives exactly the same QES point as in Section \ref{sec43}.

To calculate entropy, we evaluate the partition function in $n\rightarrow 1$ limit. It turns out that
\bal\begin{split}
\frac{\log Z_n}{n}=\frac{1}{4G_N^{(2)}}+\frac{\log Z_{n}^{\mathrm{CFT}}}{n}\ .
\end{split}\eal
Therefore, from (\ref{enp}) we can get the entropy by substituting $a=L$
\bal\begin{split}
S=\frac{1}{4G_N^{(2)}}+S_{\mathrm{matter}}([-L, L])
\end{split}\eal
which gives the QES formula used in Section \ref{sec43}.

Note that one can also achieve the extremization condition from another perspective. By varying the total action with respect to the moduli of the Riemann surface, or the position of the conical singularity $\omega$, we can also get an equation of motion, i.e.
\bal\begin{split}
0&=-\left(1-\frac{1}{n}\right) \partial_{\omega} S(\omega)+\partial_{\omega}\left(\frac{\log Z_{n}^{\mathrm{CFT}}}{n}\right)\\
&=(1-n)\partial_{\omega}\left(\frac{1}{4G_N^{(2)}}+S_{\text{matter}}([a(\omega),L])\right)\\
&=(1-n)\partial_{\omega}S_{\text{gen}}(\omega)\ .
\end{split}\eal
The last line gives the extremization of the generalized entropy.
\section{Gluing conditions}\label{gc}
In this appendix we clarify the gluing condition between the CFT on the brane and that on the asymptotic boundary for the effective boundary description.

As shown in Fig.\ref{A}, the vertical brane $Q'$ without physical degrees of freedom divides the bulk into wedges $W_1$ and $W_2$, and energy can propagate freely across $Q'$. Thus in the bulk point of view, the gluing condition is transparent. And we insist that the transparent boundary condition is maintained when switching to the boundary point of view.

 In order to satisfy the transparent boundary condition, stress tensors on the two sides should be related as \cite{Almheiri:2019yqk}
\bal
\label{str}
\left(\frac{\partial x^{+}}{\partial y^{+}}\right)^{2} T_{++}^{(x)}=T_{++}^{(y)}+\frac{c}{24 \pi}\left\{x^{+}, y^{+}\right\}\ ,
\eal
where $y^{\pm}=t\pm y$ denotes the coordinate on the brane and $y^{\pm}=t\pm y$ the coordinate on the asymtotic boundary. There is also a similar equation is for $T_{--}$ by replacing each plus sign with a minus sign. Notice that state on the asymptotic boundary is taken to be in vacuum, thus the stress tensor vanishes, i.e. $T^{(x)}_{++}=T^{(x)}_{--}=0$. And from (\ref{st}), both the left-moving and right-moving components of the stress tensor on the brane vanishes, i.e. $T^{(y)}_{++}=T^{(y)}_{--}=0$. Therefore, (\ref{str}) gives
\bal
\left\{x^{\pm}, y^{\pm}\right\}=0\ ,
\eal
which means that the coordinates $x$ and $y$ can be related to each other through an $SL(2,C)$ transformation. For our set up, the relation is simply $x^{+}=y^{+}$ and $x^{-}=y^{-}$. In other words, $x=y$ in this effective 2d boundary theory.



\end{document}